# The Generating of Fractal Images Using MathCAD Program

**Teaching Assist. Math. Laura Ştefan**
**"Dimitrie Cantemir" University of Bucharest, Romania**

ABSTACT. This paper presents the graphic representation in the z–plane of the first three iterations of the algorithm that generates the Sierpinski Gasket. It analyses the influence of the f(z) map when we represent fractal images. We considered the following maps: $f(z) = z^n, n = \overline{1,6}$ and $f(z) = \exp(z^{4/3})$.

## 1. Generating the Sierpinski Gasket

Let's begin with a triangle (for example an equilateral one). Connect the midpoints of each side to form four separate equilateral triangles and cut out the triangle in the center. For each of the three remaining triangles, perform this same act. We have

$$\lim_{k \to \infty} \bigcup_{i=1}^{3^k} T_{ki} = \bigcap_{k=1}^{\infty} \bigcup_{i=1}^{3^k} T_{ki} \qquad (1)$$

where $T_{ki}$, $i = \overline{1, 3^k}$ are triangles obtained in the k stage. We obtain a non-void, compact set named Sierpinski Gasket.

Mathcad algorithm for the graphic representation:
The iteration 1

$L := 1 \qquad j := 1..4$

$z_{1,1} := 0 + i \cdot 0 \qquad z_{1,2} := L + i \cdot 0 \qquad z_{1,3} := \frac{1}{2} \cdot L + i \cdot \frac{\sqrt{3}}{2} \cdot L \qquad z_{1,4} := z_{1,1}$

211

The iteration 2

$$z_{2,1} := \frac{1}{2} \cdot (z_{1,1} + z_{1,3}) \quad z_{2,2} := \frac{1}{2} \cdot (z_{1,1} + z_{1,2}) \quad z_{2,3} := \frac{1}{2} \cdot (z_{1,2} + z_{1,3}) \quad z_{2,4} := z_{2,1}$$

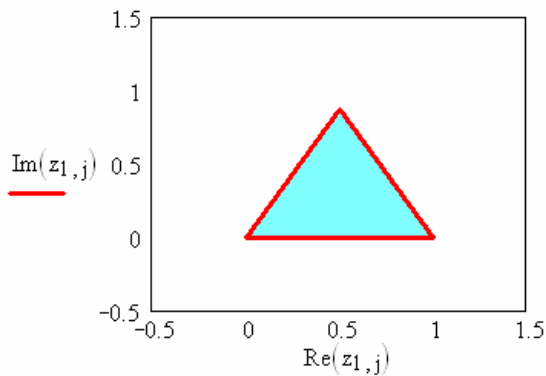

*Fig.1 The Sierpinski's triangle*

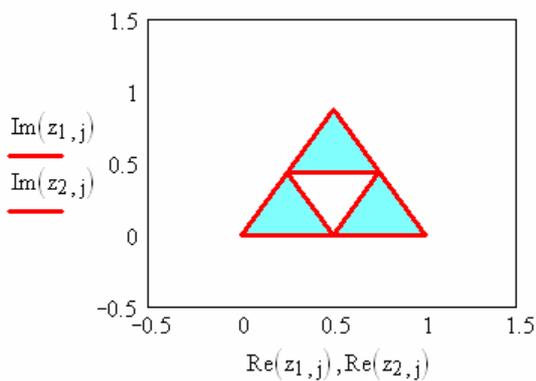

*Fig.2 The second iteration applied to Sierpinski's triangle*

The iteration 3

$$z_{31,1} := \frac{1}{2} \cdot (z_{1,1} + z_{2,1}) \quad z_{31,2} := \frac{1}{2} \cdot (z_{1,1} + z_{2,2})$$

$$z_{31,3} := \frac{1}{2} \cdot (z_{2,2} + z_{2,1}) \quad z_{31,4} := z_{31,1}$$

$$z_{32,1} := \frac{1}{2} \cdot (z_{2,2} + z_{2,3}) \quad z_{32,2} := \frac{1}{2} \cdot (z_{2,2} + z_{1,2})$$

$$z_{32,3} := \frac{1}{2} \cdot (z_{1,2} + z_{2,3}) \quad z_{32,4} := z_{32,1}$$

$$z_{33,1} := \frac{1}{2} \cdot (z_{2,1} + z_{1,3}) \quad z_{33,2} := \frac{1}{2} \cdot (z_{2,1} + z_{2,3})$$

$$z_{33,3} := \frac{1}{2} \cdot (z_{1,3} + z_{2,3}) \quad z_{33,4} := z_{33,1}$$

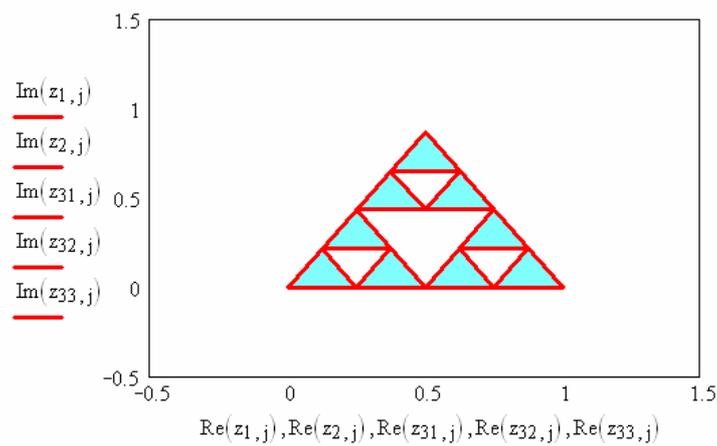

*Fig.3 The third iteration applied to Sierpinski's triangle*

## 2. The influence of the transforming function $f(z)$ in generating the fractals:

Considering a B base matrix, in Mathcad:

213

$$B = \begin{pmatrix} 0 \\ 1 \\ \frac{1}{2} + i \cdot \frac{\sqrt{3}}{2} \\ 0 \\ 100 \end{pmatrix}, \qquad (2)$$

and a second one, W, obtained from din B, as:

$$W(u1, u2, u3) = \text{augment}\left(u1^T, \text{augment}\left(u2^T, u3^T\right)\right)^T. \qquad (3)$$

The column matrix $T(B)$ is defined:

$$T(B) = \frac{1}{2} \cdot W(B, B+1, B+i), \qquad (4)$$

and the matrixes, G1, . . . G6, as:

$$G1 = B, \; G2 = T(G1), \; G3 = T(G2)$$
$$G4 = T(G3), \; G5 = T(G4), \; G6 = T(G5). \qquad (5)$$

**2.1. The case** $f1(z) = z$ **and the index** $k = 0,..5 \cdot 3^6 - 1$

Using the notation $H1 = \overrightarrow{f1(G6)}$ the H1 image is obtained in the complex plan as presented in Figure 4.

**2.2. The case** $f2(z) = z^2$ **and the index** $k = 0,..5 \cdot 3^6 - 1$

Using the notation $H2 = \overrightarrow{f2(G6)}$ the H2 image is obtained in the complex plan as presented in Figure 5.

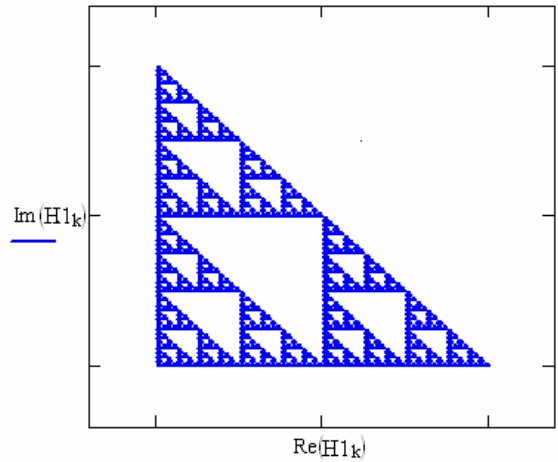

*Fig.4 The H1 map obtained using the function $f1(z) = z$ and the matrix G6*

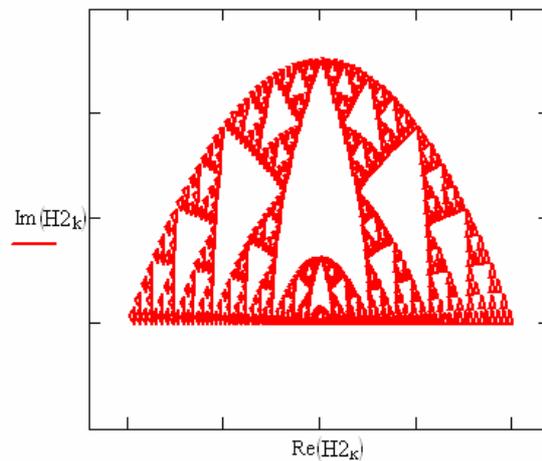

*Fig.5 The H2 map obtained using the function $f2(z) = z^2$ and the matrix G6*

**2.3. The case $f3(z) = z^3$ and the index $k = 0,..5 \cdot 3^6 - 1$**

Using the notation $H3 = \overline{f3(G6)}$ the H3 image is obtained in the complex plan:

215

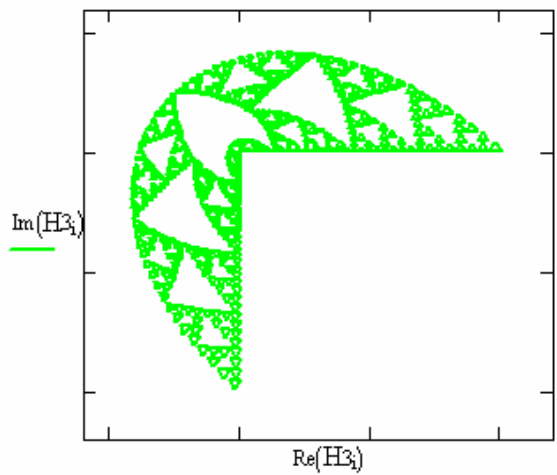

*Fig.6 The H3 map obtained using the function $f3(z) = z^3$ and the matrix G6*

**2.4. The case $f4(z) = z^4$ and index $k = 0, ... 5 \cdot 3^6 - 1$**

Using the notation $H4 = \overrightarrow{f4(G6)}$ the H4 image is obtained in the complex plan:

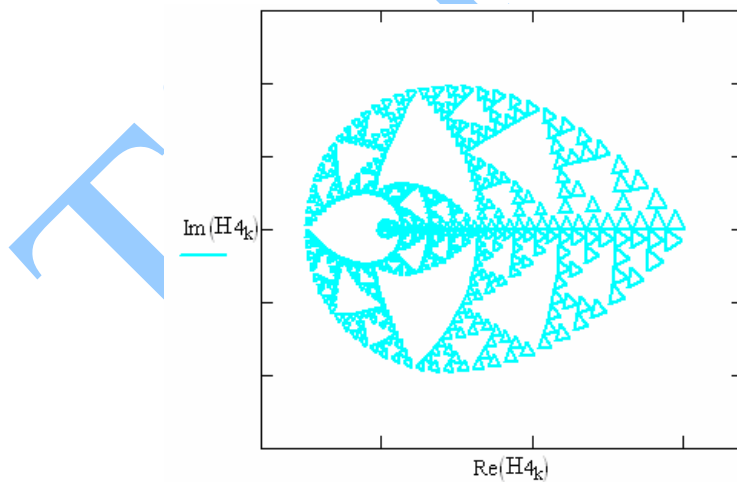

*Fig.7 The H4 map obtained using the function $f4(z) = z^4$ and the matrix G6*

**Conclusions**

The degree of the transforming function $f(z)$ produces a rotation of the image. Thus, for the first degree function the image of the fractals is represented on the first frame; for the second degree function the image of the fractals is represented on the two frames, and so on.

In the case of the sixth degree function, $f(z) = z^6$ the image of the fractals is represented in the following figure:

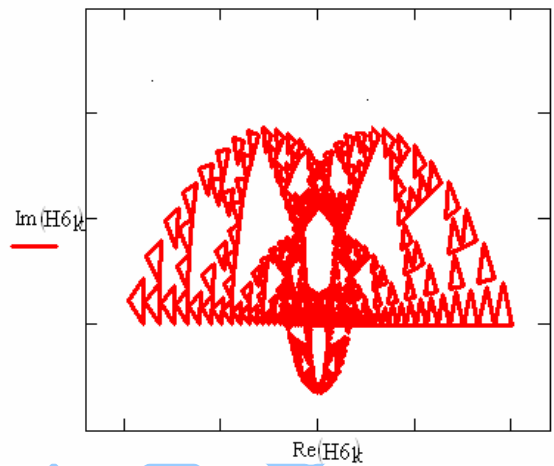

*Fig.8 The H5 image obtained using the function $f(z) = z^6$ and the matrix G6*

If instead of the G6 matrix, used in paragraph 2, one of the G3, G4 or G5 matrix is applied the spectrum of the map of the fractals is modified. Thus, for the function $f(z) = z^6$ it results:
- for G4

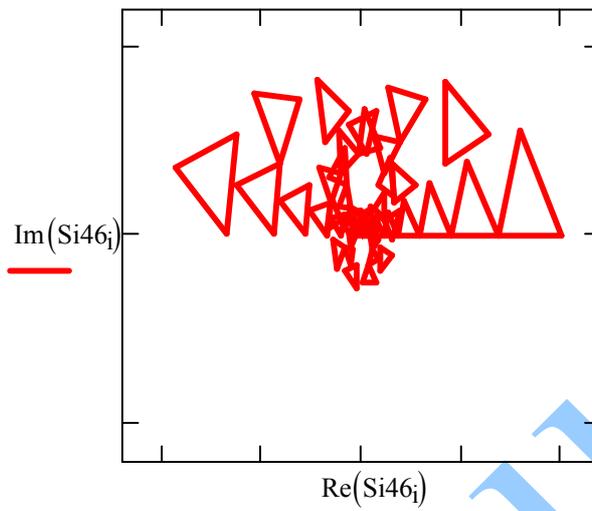

*Fig.9 The H6 map obtained using the function $f(z) = z^6$ and the matrix G4*

- for G5

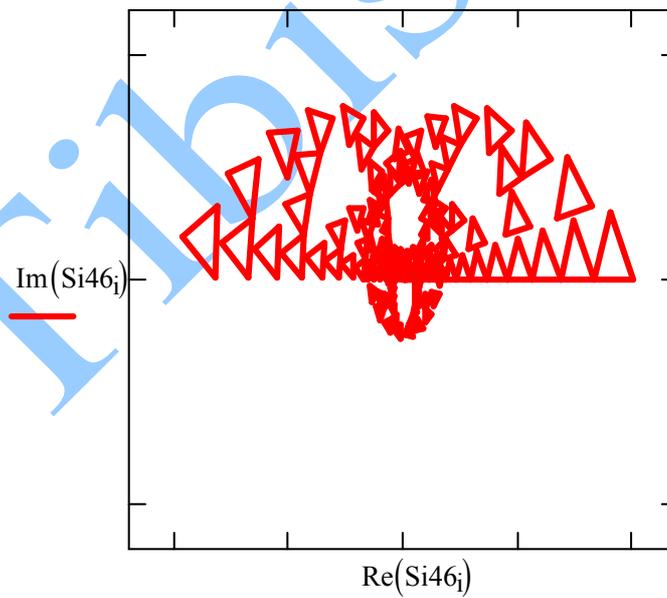

*Fig.10 The H7 map obtained using the function $f(z) = z^6$ and the matrix G5*

218

If the transforming functions are modified the domain covered by the map of the fractals is modified.

The case $f(z) = \exp(z^{4/3})$ and G6

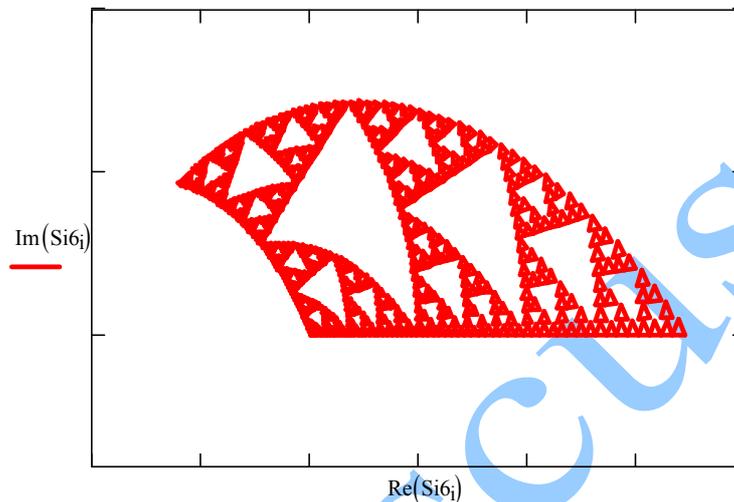

*Fig.11 The image H8 obtained using the function $f(z) = \exp(z^{4/3})$ and the matrix G6*